\documentclass[a4paper]{jpconf}
\usepackage{color}
\usepackage{amssymb}
\usepackage{amsmath}
\usepackage{amsthm} 
\usepackage{mathrsfs}                                                           
\usepackage{slashbox}  
\usepackage[all]{xy}
\usepackage{graphicx}
\usepackage{wrapfig}

\def\2{{1\over 2}}

\newcommand{\rf}[1]{(\ref{#1})}
\def\b{\bar}

\renewcommand{\t}{\tilde}
\newcommand{\p}{\partial}

\def\p{\partial}

\def\cQ{\mathcal{Q}}
\def\cE{\mathcal{E}}
\def\bcE{\bar{\mathcal{E}}}
\def\cF{\mathcal{F}}

\def\b{\bar}

\def\<{\langle}
\def\>{\rangle}
\def\+{\dagger}

\begin{document}
\title{Sigma-models and Homotopy Algebras}
\author{Anton M. Zeitlin}
\address{ Department of Mathematics, Columbia University,
\newline 2990 Broadway, New York, NY 10027, USA;
\newline
IPME RAS, \newline
V.O. Bolshoj pr., 61, 199178, St. Petersburg;\newline
zeitlin@math.columbia.edu\newline
http://math.columbia.edu/$\sim$zeitlin }

\begin{abstract}
We review the relation between homotopy algebras of conformal field theory and geometric structures arising in sigma models. In particular we formulate conformal invariance conditions, which in the quasi-classical limit are Einstein equations with extra fields, as generalized Maurer-Cartan equations. 
\end{abstract}

\section{Introduction: target space symmetries of sigma-models revisited. }
Sigma-models are among the most fruitful and puzzling objects in string theory. They can be treated as two dimensional conformal field theories (CFT), describing movement of a string in the curved background with extra fields. The corresponding conformal invariance conditions have an interpretation of nonlinear equations of motion, like Einstein equations with extra fields \cite{eeq}, \cite{fts}. 
In this article we will review some of the results of manuscripts \cite{azbg}, \cite{zeit},\cite{zeitnew}, which describe symmetries and conformal invariance conditions (nonlinear equations for the background fields) from the point of view of intrinsic homotopy  Gerstenhaber algebras of CFT. 
To do this we have to reformulate the standard classical second order sigma-model using the special first order action  \cite{lmz} and describe standard symmetries in this new language. This reformulation is given in the 
first section. Then, in section 2, the homotopy algebras of chiral CFTs are described and finally, in section 3, we formulate conjectures concerning the homotopical description of Einstein equations with B-field and dilaton in this framework. 

Let us consider a complex Riemann surface $\Sigma$, a complex manifold $M$ of dimension $d$, and a map $\rho:\Sigma\to M$. Then one can write the following first order action functional:
\begin{eqnarray}\label{free}
S_0=\frac{1}{2\pi ih}\int_\Sigma (\langle p\wedge\bar{\partial} X\rangle-
\langle \bar{p}\wedge{\partial} X\rangle),
\end{eqnarray}
where $p$ and $\bar{p}$ belong to $\rho^*(\Omega^{(1,0)}(M))\otimes \Omega^{(1,0)}(\Sigma)$ and 
$\rho^*(\Omega^{(0,1)}(M))\otimes \Omega^{(0,1)}(\Sigma)$ correspondingly, $X^i, X^{\bar j}$ stand for the pull-backs of the coordinate functions on $M$ with respect to $\rho$, and 
$\langle\cdot, \cdot\rangle$ stands for standard pairing. 
This action has the following symmetries (we write them in components in the infinitesimal form):
\begin{eqnarray}\label{set1h}
X^i\to X^i-v^i(X), \quad p_i\to p_i+\p_i v^k p_k;\quad X^{\b i}\to X^{\b i}-v^{\b i}(\b X), \quad p_{\b i}\to p_{\b i}+\p_{\b i} v^{\b k} p_{\b k} .
\end{eqnarray}
Here, the generators of the infinitesimal transformations $v, \b v$ are the elements of $\Gamma(\mathcal{O}(T^{(1,0)}M))$ and 
$\Gamma(\mathcal{O}(T^{(0,1)}M))$ correspondingly, i.e. $v^i$ ($v^{\b i}$) 
are (anti)holomorphic. These symmetries illustrate invariance under the holomorphic coordinate transformations.  
There is another set of symmetries, induced by the (anti)holomorphic 1-forms. Let 
$\omega\in \mathcal{O}({T^*}^{(1,0)}M)$ and $\b\omega\in \bar{\mathcal{O}}({T^*}^{(0,1)}M)$. Then the action \rf{free} is invariant 
under the transformation of $p, \b p$:
\begin{eqnarray}\label{set2h}
p_i\to p_i-\p X^k(\p_k\omega_i-\p_i\omega_k), \quad p_{\b i}\to p_{\b i}-\bar{\p} X^{\b k}(\p_{\b k}\omega_{\b i}-\p_{\b i}\omega_{\b k}).
\end{eqnarray}
We want to generalize the action \rf{free} so that it would be invariant under the diffeomorphism transformations and nonholomorphic generalizations of \rf{set2h}. 
In order to do that, one has to introduce extra (perturbation) terms to the action \rf{free}, so that resulting action is as follows: 
\begin{eqnarray}
\tilde{S}=
\frac{1}{2\pi ih}\int_\Sigma (\langle p\wedge\bar{\partial} X\rangle-
\langle \bar{p}\wedge{\partial} X\rangle -\langle \mu, p\wedge \b \p X\rangle -\langle \b \mu, \p X\wedge \b p\rangle-\langle b, \p X\wedge \b \p X\rangle),\nonumber
\end{eqnarray}
where $b\in \Gamma ({T^*}^{(1,0)}M\otimes {T^*}^{(0,1)}M)$.
The resulting symmetry transformations generated by $(v, \b v)$ can be written as follows:
\begin{eqnarray}\label{mutransf}
 &&\mu^{i}_{\bar{j}} \rightarrow \mu^{i}_{\bar{j}} -
\p_{\bar{j}}v^i + v^{k}\p_k\mu^{i}_{\bar{j}} +
v^{\bar{k}}\p_{\bar{k}}\mu^{i}_{\bar{j}}+
\mu^{i}_{\bar{k}}\p_{\bar{j}}v^{\bar{k}} -
\mu^k_{\bar{j}}\p_kv^i
 + \mu^i_{\bar{l}}\mu^k_{\bar{j}}\p_k v^{\bar{l}},\\ 
&&b_{i{\bar j}} \rightarrow b_{i{\bar j}} + v^k\p_k b_{i{\bar j}} + v^{\bar{k}}\p_{\bar{k}} b_{i{\bar j}}
+ b_{i{\bar k}}\p_{\bar{j}}v^{\bar{k}}+b_{l{\bar j}}\p_i v^l+b_{i{\bar k}}\mu^{k}_{\bar j}\p_kv^{\bar{k}}
+b_{l{\bar j}}{\bar\mu}^{\bar k}_i\p_{\bar k}v^l,\nonumber
 \end{eqnarray}
 and the formula for the transformation of $\bar{\mu}$ can be obtained from the one of $\mu$ by formal complex conjugation. 
This leads to the symmetry of the action $\tilde S$ if 
\begin{eqnarray}
&&X^i\to X^i-v^i(X, \b X), \quad p_{i} \rightarrow p_{i} + p_k \p_iv^k - p_k\mu^k_{\bar{l}}\p_iv^{\bar{l}}
- b_{j{\bar k}}\p_i v^{\bar k}\p X^j,\\
&& X^{\b i}\to X^{\b i}-v^{\b i}(X,\b X), \quad \bar{p}_{\bar{i}} \rightarrow \bar{p}_{\bar{i}} + \bar{p}_{\bar k} \p_{\bar i}v^{\bar {k}} - {\bar p}_{\bar{k}}
{\bar \mu}^{\bar k}_{l}\p_iv^{l}
- b_{\bar{j}k}\p_{\bar{i}} v^{k}\bar{\p} X^{\bar {j}}.\nonumber
\end{eqnarray}
Therefore, the resulting action is invariant under the action of the infinitesimal diffeomorphism group. The component formulas \rf{mutransf} were first discovered in \cite{gamayun}.  
Similarly, we obtain that the transformations 
\begin{eqnarray}\label{omtransf}
&&b_{i\bar{j}}\to  b_{i\bar{j}}+\p_{\bar j}\omega_i-\p_i\omega_{\bar j}+\mu^i_{\bar j}(\p_i\omega_k-\p_k\omega_i)+\bar{\mu}^{\bar s} _i(\p_{\bar j}\omega_{\bar s}-
\p_{\bar s}\omega_{\bar j})+
{\bar \mu}^{\bar i}_j\mu_{\bar k}^s(\p_s\omega_{\bar i}-\p_{\bar i}\omega_s)
\end{eqnarray}
accompanied with $ p_i\to p_i-\p X^k(\p_k\omega_i-\p_i\omega_k)-\p_{\b r}\omega_i\p X^{\bar r}-{\bar \mu}^{\bar s}_k\p_i\omega_{\bar s}\p X^k, ~\  p_{\b i}\to p_{\b i}-{\bar\p} X^{\b k}(\p_{\b k}\omega_{\b i}-{\p}_{\b i}\omega_{\b k})-\p_{r}\omega_{\bar i}{\bar\p} X^{r}-{\mu}^{s}_{\bar k}\p_i\omega_{s}{\bar\p} X^{\bar k} $ leave $\tilde{S}$ invariant. 
Hereinafter, it is useful to consider $\mu, {\bar \mu}, b$ as matrix elements of $\tilde{\mathbb{M}}\in \Gamma((T^{(1,0)}M\oplus {T^*}^{(1,0)}M)\otimes(T^{(0,1)}M\oplus {T^*}^{(0,1)}M)$, i.e. 
$
\tilde{\mathbb{M}}=\begin{pmatrix} 0 & \mu \\ 
\bar{\mu} & b \end{pmatrix}.
$
For simplicity of notation let us define $E=TM\oplus T^*M$, also 
$\mathcal{E}=T^{(1,0)}M\oplus {T^*}^{(1,0)}M$ and $\bar{\mathcal{E}}=T^{(0,1)}M\oplus {T^*}^{(0,1)}M$, so that $E=\mathcal{E}\oplus\bar{\mathcal{E}}$.

Let $\alpha\in \Gamma(E)$, i.e. $\alpha=(v, \bar v, \omega, \bar \omega)$, where  $v, \b v$ are the elements of $\Gamma(T^{(1,0)}M)$ and 
$\Gamma(T^{(0,1)}M)$ correspondingly 
and $\omega\in \Omega^{(1,0)}(M)$, $\b\omega\in \Omega^{(0,1)}(M)$. Next, we introduce an operator 
$D:\Gamma(E)\to \Gamma(\cE\otimes\bcE)$, such that
$D\alpha=\left( \begin{array}{cc}
0 & {\bar \p }v\\
{\p \bar v} & \p{\bar\omega}-{\bar\p} \omega \end{array} \right)$.
Then the transformation of $\tilde{\mathbb{M}}$ under \rf{mutransf}, \rf{omtransf} can be expressed by the following formula:
\begin{eqnarray}\label{algsym}
\tilde{\mathbb{M}}\to \tilde{\mathbb{M}}-D\alpha+ \phi_1(\alpha,\tilde{\mathbb{M}})+\phi_2(\alpha, \tilde{\mathbb{M}},\tilde{\mathbb{M}}).
\end{eqnarray}
The second operation $\phi_1(\alpha,\tilde{\mathbb{M}})$ can be described as follows. Let us consider 
\begin{eqnarray}\label{jets}
\xi\in J^{\infty}(\mathcal{O}_M)\otimes 
J^{\infty}({\bar{\mathcal{O}}}(\bcE))
\oplus J^{\infty}({\mathcal{O}}(\cE))
\otimes J^{\infty}({\bar{\mathcal{O}}}_M),\quad \mathbb{L}\in J^{\infty}({\mathcal{O}}(\cE))\otimes J^{\infty}({\bar{\mathcal{O}}}(\bcE)),
\end{eqnarray}
where $J^{\infty}(E)$ for any bundle $E$ over $M$ stands for the corresponding $\infty$-jet bundle of $E$.  
In other words, let $
\xi=\sum_Jf^J\otimes {\bar{b}}^J+\sum_Kb^K\otimes {\bar{f}}^K,\quad \mathbb{L}=\sum_I a^I\otimes \bar{a}^I, $
where $a^I, b^J\in J^{\infty}({\mathcal{O}}(\cE))$, $f^I\in 
J^{\infty}(\mathcal{O}_M)$ and ${\bar{a}}^I, {\bar{b}}^J\in J^{\infty}({\bar{\mathcal{O}}}(\bcE))$, 
${\bar{f}}^I\in 
J^{\infty}(\bar{\mathcal{O}}_M)$. Then we can introduce an operation $\phi_1(\xi, \mathbb{L})$ as follows:
\begin{eqnarray}\label{bialgop}
\phi_1(\xi,\mathbb{L})=
\sum_{I,J}[b^J, a^I]_D\otimes 
{\bar{f}}^{J}{\bar{a}}^I+\sum_{I,K}f^Ka^I\otimes[{\bar{b}}^K, {\bar{a}}^I]_D,
\end{eqnarray}
where $[\cdot, \cdot]_D$ is a Dorfman bracket (see the definition in  the next section). Completing the tensor products 
in \rf{jets}, we find that the operation $\phi_1$ can be induced on $\alpha\in \Gamma(E)$ and 
$\tilde{\mathbb{M}}\in\Gamma(\cE\otimes \bcE)$. 
One can explicitly check that 
\rf{bialgop} leads to the part of \rf{mutransf} and \rf{omtransf}, linear in $\alpha$ and $\tilde{\mathbb{M}}$. The last part, bilinear in 
$\tilde{\mathbb{M}}$, also has an algebraic meaning of a similar kind: returning back to the jet notation, we find that on the jet counterparts 
of $\alpha, \tilde{\mathbb{M}}$, i.e. on $\xi, \mathbb{L}$ the expression for $\phi_2$ is:
\begin{eqnarray}\label{trialgop}
\phi_2(\xi, \mathbb{L},\mathbb{L})=
\frac{1}{2}\sum_{I,J,K}\langle b^I, a^K\rangle a^J\otimes \bar{a}^J(\bar{f}^I) {\bar{a}}^K+\frac{1}{2}\sum_{I,J,K}
 a^J(f^I) {a}^K\otimes \langle {\bar{b}}^I, {\bar{a}}^K\rangle {\bar{a}}^J,
\end{eqnarray}
where $\bar{a}^J(\bar{f}^I)$, $a^J(f^I)$ correspond to the action of the differential operator, associated to the vector field, on a function 
($\bar{a}^J(\bar{f}^I)$, $a^J(f^I)$ are set to be zero if $\bar{a}^J$, 
$a^J$ are 1-forms).  
At the same time, the operation $\phi_2$ 
has the following simple description:
$
\phi_2(\alpha, \tilde{\mathbb{M}},\tilde{\mathbb{M}})=\tilde{\mathbb{M}}\cdot D\alpha\cdot \tilde{\mathbb{M}}
$
if we consider $\tilde{\mathbb{M}}$ as an element of $End(\Gamma(E))$.

Let us notice that we could generalize $\tilde{\mathbb{M}}$ in the following way: in the matrix expression for 
$\tilde{\mathbb{M}}$ let us fill in the empty spot, i.e. let us add extra element $g\in\Gamma(T^{(1,0)}M\otimes T^{(0,1)}M)$. Then the modified 
$\tilde{\mathbb{M}}$, i.e.    
$\mathbb{M}\in\Gamma(\cE\otimes\bcE)$ can be expressed as follows: 
$
\mathbb{M}=\begin{pmatrix} g & \mu \\ 
\bar{\mu} & b \end{pmatrix}.
$
The corresponding first order action functional is:
\begin{eqnarray}
&&S_{fo}=\\
&&\frac{1}{2\pi i h}\int_\Sigma (\langle p\wedge\bar{\partial} X\rangle-
\langle \bar{p}\wedge{\partial} X\rangle -\langle g, p\wedge \bar{p} \rangle-\langle \mu, p\wedge \b \p X\rangle -\langle \b \mu, \b p\wedge \p X\rangle-\langle b, \p X\wedge \b \p X\rangle).\nonumber
\end{eqnarray}
It turns out that the symmetries of this action functional can be described by the same formula \rf{algsym}, where the algebraic meaning of 
operations on the jet level is given by the same formulas \rf{bialgop}, \rf{trialgop}. One can easily reproduce explicit component formulas for the infinitesimal symmetries of the action $S_{fo}$.
The reason for introducing the $g$-term in the action functional is as follows. If the matrix $\{g^{i\bar{j}}\}$ is invertible, 
then using elementary variational calculus, one can find that 
the critical points for $S_{fo}$ are the same as for the 
second-order action functional:
\begin{eqnarray}
S_{so}=\frac{1}{4\pi h}\int_{\Sigma} d^2 z
(G_{\mu\nu}+B_{\mu\nu})\partial X^{\mu}\bar{\partial}X^{\nu},
\end{eqnarray}
where $G$ is a symmetric tensor and $B$ is antisymmetric one, indices $\mu,\nu$ run through the set $\{i, \bar{j}\}$. The expression for 
$G$ and $B$ via $\mathbb{M}$ is given by:
\begin{eqnarray}\label{GB}
\label{phytwi}
G_{s\bar{k}}&=&g_{\bar{i}j}
\bar{\mu}^{\bar{i}}_s\mu^{j}_{\bar{k}}+g_{s\bar{k}}-
b_{s\bar{k}}, \quad
B_{s\bar{k}}=g_{\bar{i}j}\bar{\mu}^{\bar{i}}_s\mu^{j}_{\bar{k}}-g_{s\bar{k}}-
b_{s\bar{k}},\\
G_{si}&=&-g_{i\bar{j}}\bar{\mu}^{\bar{j}}_s-g_{s\bar{j}}\bar{\mu}^{\bar{j}}_i
, \quad
G_{\bar{s}\bar{i}}=-g_{\bar{s}j}\mu^{j}_{\bar{i}}-g_{\bar{i}j}\mu^{j}_{\bar{s}},
\nonumber\\
B_{si}&=&g_{s\bar{j}}\bar{\mu}^{\bar{j}}_i-g_{i\bar{j}}\bar{\mu}^{\bar{j}}_s,
\quad
B_{\bar{s}\bar{i}}=g_{\bar{i}j}\mu^{j}_{\bar{s}}-g_{\bar{s}j}\mu^{j}_{\bar{i}},
\nonumber
\end{eqnarray}
where $\{g_{i\bar{j}}\}$ stands for the inverse matrix of $\{g^{i\bar{j}}\}$. Such parametrization of the second-order action in the case when $M$ is a Riemann surface was first introduced in \cite{pz}, \cite{zeitlin}.
The symmetries of the action functional $S_{fo}$ transform 
into infinitesimal diffeomorphism transformations and the 2-form $B$ symmetry
$ G\to G-L_{\bf v}G,$ $B\to B-L_{\bf v}B$; $ B\to B-2d{\bf\boldsymbol \omega},$
 if $\alpha=({\bf v}, {\boldsymbol \omega})$, so that ${\bf v}\in \Gamma(TM)$, ${\boldsymbol \omega}\in \Omega^{1}(M)$, i.e. the symmetries of $S_{so}$. 

Let us formulate this as a theorem.\\

\noindent {\bf Theorem 1.1.} {\it Let $\mathbb{M}\in\Gamma(\cE\otimes\bcE)$, parametrized as above, so that its $\Gamma(T^{(1,0)}M\otimes T^{(0,1)}M)$ part is given by $\{g^{i\bar{j}}\}$, which is invertible, 
then the infinitesimal diffeomorphism transformations of the resulting symmetric and antisymmetric tensors $G$ and $B$ (see \rf{GB}), as well as the 
B-tensor shift by exact 2-forms are encoded in the formula $
\mathbb{M}\to \mathbb{M}-D\alpha+ \phi_1(\alpha,\mathbb{M})+\phi_2(\alpha, \mathbb{M},\mathbb{M})$,  
where $\alpha\in \Gamma(E)$ and operations $\phi_1,\phi_2$ are defined above.}\\

Note that if $\{G_{\mu\nu}\}$ is invertible and real, it gives rise to the metric tensor. Therefore, since $\mathbb{M}$ parametrizes both 
$G$ and $B$, and transforms according to \rf{algsym} under diffeomorphisms, it is analogous to Beltrami differential on the  Riemann surface. 
So, from now on we will call the elements of $\Gamma(\cE\otimes \bcE)$ as Beltrami-Courant differentials, since, as we see in the following sections, they are described by means of the Courant algebroid \cite{courant} structure on $\mathcal{E}, \bcE$. 

\section{Vertex algebroids, $G_{\infty}$-algebra and quasiclassical limit}
In this section, we describe the constructions of the article \cite{zeit} with some modifications and refer the reader to this article for some of the details. Namely, we discuss homotopy algebras of 
$G_{\infty}$- and $BV_{\infty}$- type (see e.g. \cite{tamarkin}), emerging from the vertex algebras associated with semi-infinite complex (see e.g. \cite{lz}).

Each of the terms in the classical action $S_0$ 
from which we started the previous section, leads to the quantum theory which is well described locally on open neighborhoods of $M$ by means of vertex algebra generated by operator products 
$X^i(z)p_j(w)\sim \frac{h\delta^i_j}{z-w}, \quad 
X^{\b i}(\b z)p_{\b j}(\b w)\sim \frac{h\delta^{\b i}_{\b j}}{\b z-\b w}$ and globally by means of gerbes of chiral differential operators on $M$  \cite{gms}. 
Each of the corresponding vertex algebras, which provide the local description, 
form a $\mathbb{Z}_{+}$-graded vector space $V=\sum_{n=0}^{+\infty}V_n$, so that it is determined (see \cite{gms}) 
by means of a vertex algebroid. 
In our case, the vertex algebroid is described by means of the sheaf 
$\mathcal{V}=\mathcal{O}(\cE)\otimes \mathbb{C}[h]\equiv \mathcal{O}(\cE)^h$ (resp. $\bar{\mathcal{O}}(\bcE)^h$), of vector spaces $V_1$, as well as the sheaf of $V_0$ spaces, which coincides with the structure sheaf $\mathcal{O}_M\otimes \mathbb{C}[h]=\mathcal{O}_M^h$ (resp. $\bar{\mathcal{O}}_M^h$), with certain algebraic operations between them.
 
Let us define a vertex algebroid (see e.g. \cite{gms}, \cite{bressler}) and then study our concrete case in detail. 

A {\em vertex $\mathcal{O}_M$-algebroid} is a sheaf of $\mathbb{C}$-vector 
spaces $\mathcal{V}$ with a pairing $
\mathcal{O}_M\otimes_{\mathbb{C}[h]}\mathcal{V}  \to  \mathcal{V}$, i.e.  
$f\otimes v  \mapsto  f*v$ 
such that $1* v = v$, equipped with
a structure of a Leibniz $\mathbb{C}[h]$-algebra 
$[\ ,\ ] :
\mathcal{V}\otimes_{\mathbb{C}[h]}\mathcal{V}\to \mathcal{V}$, a $\mathbb{C}[h]$-linear map of Leibniz algebras $\pi : \mathcal{V}\to \Gamma({TM})$, which 
usually is referred to as an anchor, 
a symmetric $\mathbb{C}[h]$-bilinear pairing $\langle\ ,\ \rangle :
\mathcal{V}\otimes_{\mathbb{C}[h]}\mathcal{V}\to \mathcal{O}_M^h$
a $\mathbb{C}$-linear map $\p : \mathcal{O}_M\to \mathcal{V}$ 
such that
$\pi\circ\partial = 0$,
which satisfy certain relations identical to the ones satisfied between $V_1$ and $V_0$ subspaces of the nonnegatively graded vertex algebra $V$. 

Let us concentrate on the case when 
$\mathcal{V}=\mathcal{O}(\cE)^h$. Explicitly, if 
$f\in \mathcal{O}_M$, $v, v_1, v_2\in \mathcal{O}(T^{(1,0)}M)$, 
$\omega, \omega_1, \omega_2\in \mathcal{O}({T^*}^{(1,0)}M)$, then 
locally in the neighborhood with the coordinates $\{X^i\}$
\begin{eqnarray}\label{va}
&&\p f=df,\quad \pi(v)f=-hv(f), \quad\pi(\omega)=0,\quad f*v=fv+hdX^i\p_i\p_jfv^j, \quad f*\omega=f\omega,\nonumber\\
&&[v_1,v_2]=-h[v_1,v_2]_D-h^2dX^i\p_i\p_kv^s_1\p_sv^k_2,\quad [v,\omega]=-h[v,\omega]_D, \quad 
[\omega, v]=-h[\omega,v]_D, \nonumber\\
&& [\omega_1,\omega_2]=0,\quad \langle v, \omega\rangle=-h\langle v, \omega\rangle^s, \quad \langle v_1, v_2\rangle=-h^2\p_iv_1^j\p_jv_2^i,\quad \langle\omega_1, \omega_2\rangle=0,
\end{eqnarray}
where $\langle\cdot,\cdot \rangle^s$ is a standard pairing on $\cE$ and $[\cdot,\cdot]_D$ is the Dorfman bracket:
\begin{eqnarray}
&&[v_1,v_2]_D=[v_1,v_2]^{Lie}, \quad [v,\omega]_D=L_v\omega,\quad [\omega , v]_D=-i_vd\omega,\quad [\omega_1,\omega_2]_D=0.
\end{eqnarray}
In \cite{zeit}, it was shown that given a holomorphic volume form on the open neighborhood $U$ of $M$, one can associate a homotopy Gerstenhaber algebra to the vertex algebroid on $U$  
(although the main emphasis of \cite{zeit} was on $C_{\infty}$ part of 
it). 
This was done by considering semi-infinite complex associated to the vertex algebra \cite{fgz}: due to the results of\cite{lz}, \cite{kvz}, \cite{huangzhao}, \cite{voronov}, there is a structure of $G_{\infty}$ algebra attached to it if the  central charge of the corresponding Virasoro algebra is 26. Using this fact and considering the subcomplex corresponding to the elements of total conformal weight zero, we find out that the central charge condition can be dropped. 
The resulting complex $(\mathcal{F}^{\cdot}, Q)$ appears to be much shorter than original semi-infinite one:  
\begin{eqnarray}
\xymatrixcolsep{30pt}
\xymatrixrowsep{3pt}
\xymatrix{
& \mathcal{V}\ar[ddddr] & \mathcal{V}\ar[ddddr]^{\frac{1}{2}h{\rm div}}& \\
&& \p &&\\
& \bigoplus & \bigoplus & \\
&& {-\frac{1}{2}h{\rm div}} &&\\
\mathcal{O}_M^{h}\ar[uuuur]^{\p} & \mathcal{O}_M^{h}\ar[uuuur]\ar[r]_{i\rm{d}} & \mathcal{O}_M^{h}  & \mathcal{O}_M^{h}.
}
\end{eqnarray}
Here $\mathcal{F}_h^0\cong\mathcal{O}_M^{h}\cong\mathcal{F}_h^3$, $\mathcal{F}_h^1\cong\mathcal{O}_M^{h}\oplus \mathcal{V}\cong\mathcal{F}_h^3$, and ${\rm div}$ stands for  
divergence operator with respect to the nonvanishing volume form applied to sections of $\Gamma(U,T^{(1,0)}(M))$. Appropriate analogue of operator ${\rm div}$ in the case of general vertex algebroid is called $Calabi-Yau$ $structure$ on vertex algebroid \cite{gms} (since e.g. in our case to be defined globally $M$ should possess a nonvanishing holomorphic volume form).  
According to \cite{zeit}, this complex has a bilinear operation $(\cdot, \cdot)_h$, which satisfies the Leibniz identity with respect to $Q$, it 
is also homotopy commutative and associative, and can be described 
via the operations from vertex algebroid. 

We note that there is an operator $\mathbf{b}$ of degree -1 on 
$(\mathcal{F}_h^{\cdot}, Q)$ which anticommutes with $Q$:  
\begin{eqnarray}
\xymatrixcolsep{30pt}
\xymatrixrowsep{3pt}
\xymatrix{
& \mathcal{V} & \mathcal{V}\ar[l]_{-i\rm{d}}& \\
& \bigoplus & \bigoplus & \\
\mathcal{O}_M^{h} & \mathcal{O}_M^{h}\ar[l]_{i\rm{d}} & \mathcal{O}_M^{h}& \mathcal{O}_M^{h}\ar[l]_{-i\rm{d}}
} 
\end{eqnarray}
This operator gives rise to the bracket operation 
\begin{eqnarray}\label{brack}
(-1)^{|a_1|}\{a_1,a_2\}_h=\mathbf{b}(a_1,a_2)_h-(\mathbf{b}a_1,a_2)_h-(-1)^{|a_1|}(a_1,\mathbf{b}a_2)_h,
\end{eqnarray}
satisfying quadratic relations together with $(\cdot, \cdot)_h$ and $Q$, which follow from the properties of vertex algebra \cite{lz}. On the cohomology of $Q$ these relations turn 
into the defining properties of Gerstenhaber algebra. Namely, the following Proposition holds.\\

\noindent {\bf Proposition 2.1.}\cite{zeit} {\it Symmetrized versions of operations $(\cdot, \cdot)_h $ together with \rf{brack} satisfy the relations of the homotopy Gerstenhaber algebra, which follow from these relations: 
\begin{eqnarray}\label{lzrel}
&&Q(a_1,a_2)_h=(Q a_1,a_2)_h+(-1)^{|a_1|}(a_1,Q a_2)_h,\\
&&(a_1,a_2)_h-(-1)^{|a_1||a_2|}(a_2,a_1)_h=Qm_h(a_1,a_2)+m_h(Qa_1,a_2)+(-1)^{|a_1|}m_h(a_1,Qa_2),\nonumber\\
&& Q(a_1,a_2,a_3)_h+(Qa_1,a_2,a_3)_h+(-1)^{|a_1|}(a_1,Qa_2,a_3)_h+\nonumber\\
&&(-1)^{|a_1|+|a_2|}(a_1,a_2,Qa_3)_h=((a_1,a_2)_h,a_3)_h-(a_1,(a_2,a_3)_h)_h,\nonumber\\
&&\{a_1,a_2\}+(-1)^{(|a_1|-1)(|a_2|-1)}\{a_2,a_1\}=\nonumber\\
&&(-1)^{|a_1|-1}(Qm_h'(a_1,a_2)-m_h'(Qa_1,a_2)-(-1)^{|a_2|}m_h'(a_1,Qa_2)),
\nonumber\\
&& \{a_1,(a_2,a_3)_h\}_h=(\{a_1,a_2\}_h,a_3)_h+(-1)^{(|a_1|-1)||a_2|}(a_2,\{a_1, a_3\}_h)_h,\nonumber\\
&&\{(a_1,a_2)_h,a_3\}_h-(a_1,\{a_2,a_3\}_h)_h-(-1)^{(|a_3|-1)|a_2|}(\{a_1,a_3\}_h,a_2)_h=\nonumber\\
&&(-1)^{|a_1|+|a_2|-1}(Qn_h'(a_1,a_2,a_3)-n_h'(Qa_1,a_2,a_3)-\nonumber\\
&&(-1)^{|a_1|}n_h'(a_1,Qa_2,a_3)-(-1)^{|a_1|+|a_2|}n_h'(a_1,a_2,Qa_3),\nonumber\\
&&\{\{a_1,a_2\}_h,a_3\}_h-\{a_1,\{a_2,a_3\}_h\}_h+(-1)^{(|a_1|-1)(|a_2|-1)}\{a_2,\{a_1,a_3\}_h\}_h=0,\nonumber
\end{eqnarray}
where $m_h,m'_h$ are some bilinear operations of degrees $-1$, $-2$ correspondingly and $n_h, n'_h$ are trilinear operations of degree -1, -2 correspondingly. There exist higher homotopies which turn this homotopy Gerstenhaber algebra into $G_{\infty}$ algebra.}\\

The last part of Proposition 2.1 follows from the results of \cite{huangzhao}, \cite{voronov}, \cite{kvz} where it was show that the symmetrized versions of $(\cdot, \cdot)_h$, $\{,\}_h$ can be continued to the $G_{\infty}$ algebra \cite{tamarkin}. One should 
also consult \cite{gorbounov}, \cite{vallette} for a different proof in the case of positively graded topological vertex algebras.

One of the central observations of \cite{zeit} was that this  $G_{\infty}$ algebra has  $quasiclassical$ limit, which can be constructed as follows.
Let $\mathcal{V}\vert_{h=0}=\mathcal{V}^0$ 
(in our example $\mathcal{V}^0=\mathcal{O}(\mathcal{E})$), then consider the subcomplex of  $(\mathcal{F}_h^{\cdot}, Q)$, i.e.  $(\mathcal{F}^{\cdot}, Q)\cong (\mathcal{F}_1^{\cdot}, Q)$, which is:
\begin{eqnarray}
\xymatrixcolsep{30pt}
\xymatrixrowsep{3pt}
\xymatrix{
& \mathcal{V}^0\ar[ddddr] & h\mathcal{V}^0\ar[ddddr]^{\frac{1}{2}h{\rm div}}& \\
&& \p &&\\
& \bigoplus & \bigoplus & \\
&& {-\frac{1}{2}h{\rm div}} &&\\
\mathcal{O}_M\ar[uuuur]^{\p} & h\mathcal{O}_M\ar[uuuur]\ar[r]_{i\rm{d}} & h\mathcal{O}_M  & h^2\mathcal{O}_M
}
\end{eqnarray}
It is easy to see that
\begin{eqnarray}
(\cdot,\cdot)_h: \cF^i\otimes \cF^j\to \cF^{i+j}[h], \quad  \{\cdot,\cdot\}_h: \cF^i\otimes \cF^j\to h\cF^{i+j-1}[h],~\ {\bf b}: \cF^i\to h\cF^{i-1}[h], 
\end{eqnarray}
so that $
(\cdot,\cdot)_0= \lim_{h\to 0}(\cdot,\cdot)_h, ~\ 
\{\cdot,\cdot\}_0= \lim_{h\to 0}h^{-1}\{\cdot,\cdot\}_h, ~\ \mathbf{b}_0=\lim_{h\to 0}h^{-1}\mathbf{b}$
are well defined. The corresponding homotopy Gerstenhaber algebra is 
much less complicated: the corresponding $L_{\infty}$ and $C_{\infty}$ parts are only $L_3$ and $C_3$-algebras. Let us have a look in detail. 
On the level of the vertex algebroid of $\mathcal{O}({\cE})^h$, let us  denote $\lim_{h\to 0}{h}^{-1}[v_1,v_2]=[v_1,v_2]_0, ~\
\lim_{h\to 0}{h}^{-1}\pi=\pi_0, ~\ \lim_{h\to 0}{h}^{-1}\langle\cdot, \cdot \rangle= \langle\cdot, \cdot \rangle_0.$
Therefore, we can express the bilinear operations $(\cdot, \cdot)_0$ and $\{\cdot,\cdot\}_0$  on the  complex 
\begin{eqnarray}
\xymatrixcolsep{30pt}
\xymatrixrowsep{3pt}
\xymatrix{
& \mathcal{O}(\mathcal{E})\ar[ddddr] & \mathcal{O}(\mathcal{E})\ar[ddddr]^{\frac{1}{2}{\rm div}}& \\
&& d &&\\
& \bigoplus & \bigoplus & \\
&& {-\frac{1}{2}{\rm div}} &&\\
\mathcal{O}_M\ar[uuuur]^{d} & \mathcal{O}_M\ar[uuuur]\ar[r]_{i\rm{d}} & \mathcal{O}_M  & \mathcal{O}_M
}
\end{eqnarray}
completely in terms of the operations of pure covariant parts of \rf{va} (explicitly you can find it in \cite{zeit}).
Let us summarize results about the quasi-classical limit via  proposition.\\

\noindent{\bf Proposition 2.2.}\cite{zeit} {\it The operations $(\cdot, \cdot)_0$, $\{\cdot,\cdot\}_0$ satisfy the relations \rf{lzrel} so that their symmetrized versions satisfy the relations of $G_{\infty}$ algebra which is the quasiclassical limit of $G_{\infty}$ algebra considered in Proposition 2.1. 
The resulting $C_{\infty}$ and $L_{\infty}$ algebras are reduced to $C_3$ and $L_3$ algebras.  
}\\

The classical limits for the corresponding homotopies $m_h=m_0+O(h)$ and $n_h=n_0+O(h)$ are as follows. The commutativity homotopy $m_0$ is nonzero iff its both arguments belong to $\cF_1$: $
m_0=-\langle A_1, A_2\rangle_0. 
$
The associativity homotopy $n_0$ is nonzero only when all three elements belong to $\cF_1$ or one of the first 
two belongs to $\cF_2$ and the other belongs to $\cF_1$:
\begin{eqnarray}
&& n_0(A_1,A_2,A_3)= A_2\langle A_1,A_3\rangle_0-A_1\langle A_2,A_3\rangle_0,~\  n_0(A_1,\t v,A_2)=n_0(\t v,A_1, A_2)=-\t v \langle A_1, A_2\rangle_0.\nonumber
\end{eqnarray}

As we stated before in the quasiclassical limit we get rid of all noncovariant terms in the expression for the product and the bracket: this is very close to the classical limit procedure for vertex algebroid. Namely, using our limit procedure, one can obtain Courant algebroid from vertex algebroid. 

The definition of Courant algebroid is as follows (see e.g. \cite{courant}, \cite{bressler}).
A  Courant $\mathcal{O}_M$-algebroid is an $\mathcal{O}_M$-module $\cQ$
equipped with the structure of a Leibniz $\mathbb{C}$-algebra
$[ , ]_0 : \cQ\otimes_\mathbb{C}\cQ \to \cQ $, 
an $\mathcal{O}_M$-linear map of Leibniz algebras (the anchor map)
$
\pi_0 : \cQ \to \Gamma(TM)
$,
a symmetric $\mathcal{O}_M$-bilinear pairing
$\langle\cdot, \cdot\rangle: \cQ\otimes_{\mathcal{O}_M}\cQ \to \mathcal{O}_M $,  
a derivation
$
\p : \mathcal{O}_M \to \cQ
$,
which satisfy
\begin{eqnarray}
&&\pi\circ\partial =  0, \quad[q_1,fq_2]_0 = f[q_1,q_2] + \pi_0(q_1)(f)q_2,~\
\langle [q,q_1],q_2\rangle + \langle q_1,[q,q_2]\rangle  =  \pi_0(q)(\langle q_1, q_2\rangle_0), \nonumber\\
&& [q,\partial(f)]_0  =  \partial(\pi_0(q)(f)), ~\ \langle q,\partial(f)\rangle  =  \pi_0(q)(f) \quad [q_1,q_2]_0 + [q_2,q_1]_0  =  \partial(\langle q_1, q_2\rangle_0), 
\end{eqnarray}
where $f\in\mathcal{O}_M$ and $q,q_1,q_2\in\cQ$. In our case $\cQ\cong\mathcal{O}(\cE)$, $\pi_0$ is just a projection on $\mathcal{O}(TM)$ and 
$[q_1,q_2]_0=-[q_1,q_2]_D, ~\ \langle q_1, q_2\rangle_0=-\langle q_1, q_2\rangle^s, ~\ \p=d.$

As we indicated earlier, both $C_{\infty}$ and $L_{\infty}$  parts of $G_{\infty}$ algebra appear to be short. We expect this to happen with all the homotopies, i.e. it is natural to suggest that $G_{\infty}$- algebra of Proposition 2.2 has only bilinear and trilinear operations, i.e. it is a $G_3$ algebra.
In the following, since we are interested only in the quasiclassical algebra on the complex $(\mathcal{F}^{\cdot},Q)$, we will neglect the 0 subscript for all multilinear operations of this algebra. 

\section{Homotopy Gerstenhaber algebra and Einstein equations}

\noindent{\bf 3.1. Nontrivial example. }
The homotopy Gerstenhaber algebra we studied in the previous section,  has a subalgebra based on the following complex $(\mathcal{F}^{\cdot}_{sm}, \mathcal{Q})$. 
\begin{eqnarray}
\xymatrixcolsep{30pt}
\xymatrixrowsep{3pt}
\xymatrix{
& \mathcal{O}(T^{(1,0)}M)\ar[ddddr] & \mathcal{O}(T^{(1,0)}M)\ar[ddddr]^{\frac{1}{2}{\rm div}}& \\
&& 0 &&\\
& \bigoplus & \bigoplus & \\
&& {-\frac{1}{2}{\rm div}} &&\\
\mathbb{C}\ar[uuuur]^{0} & \mathbb{C}\ar[uuuur]\ar[r]_{i} & \mathcal{O}_M  & \mathcal{O}_M
}
\end{eqnarray}
It is just a Gerstenhaber algebra (with no higher homotopies), moreover it is a BV algebra \cite{tamarkin}, since ${\mathbf b}$ operator also preserves $(\mathcal{F}^{\cdot}_{sm}, Q)$. Therefore, we have the following proposition.\\

\noindent{\bf Proposition 3.1.} {\it Bilinear operations $(\cdot,\cdot)$, $\{\cdot,\cdot\}$ together  with operator $\mathbf{b}$ generate the structure of BV algebra on $(\mathcal{F}^{\cdot}_{sm}, Q)$.}\\
 
Let us consider the ${\infty}$-jet version of the complex $(\mathcal{F}^{\cdot}_{sm}, Q)$: we substitute $\mathcal{O}_M$, $\mathcal{O}(T^{(0,1)}(M))$ by $J^{\infty}(\mathcal{O}_M)$, $J^{\infty}(\mathcal{O}(T^{(0,1)}(M))$. We denote the  resulting complex as $(\mathcal{F}^{\cdot}_{sm, \infty}, Q)$. 
Then let us introduce a completed tensor product ${\bf F}^{\cdot}_{sm, \infty}=\mathcal{F}^{\cdot}_{sm,\infty}\hat{\otimes}\bar {\mathcal{F}}^{\cdot}_{sm, \infty},$ 
where $(\bar{\mathcal{F}}^{\cdot}_{sm, \infty}, \bar{Q})$ is the  complex obtained from $(\mathcal{F}^{\cdot}_{sm, \infty}, Q)$ by complex conjugation. Complex 
$({\bf F}^{\cdot}_{sm, \infty}, \mathcal{Q})$, where $\mathcal{Q}=Q+\bar{Q}$, 
is the jet version of the complex $({\bf F}_{sm}^{\cdot}, \mathcal{Q})$, such 
that e.g. ${\bf F}_{sm}^2=\Gamma(T^{(1,0)}M\otimes T^{(0,1)}M)\oplus \bar{\mathcal{O}}(T^{(0,1)}M)\oplus \mathcal{O}(T^{(1,0)}M)\oplus
\mathcal{O}_M\oplus \bar{\mathcal{O}}_M\oplus \mathbb{C}$. 
Clearly, the complex $({\bf F}_{sm}^{\cdot}, \mathcal{Q})$ carries the 
structure of BV algebra inherited from $(\mathcal{F}^{\cdot}_{sm, \infty}, {Q})$ and its complex conjugation, so that 
\begin{eqnarray}\label{brack2}
(-1)^{|a_1|}\{a_1,a_2\}=\mathbf{b^-}(a_1,a_2)-(\mathbf{b^-}a_1,a_2)-(-1)^{|a_1|}(a_1\mathbf{b^-}a_2)\nonumber,
\end{eqnarray}
where $\mathbf{b^-}=\mathbf{b}-\bar{\mathbf{b}}$. Note, that the elements closed under $\mathbf{b}^-$ form a subalgebra in the differential graded Lie algebra (DGLA), generated by $Q, \{\cdot, \cdot\}$. 
It turns out that the Maurer-Cartan equations of this DGLA and their symmetries have a very interesting meaning. To describe them, let us define some extra algebraic operations for convenience.

Let $g,h \in \Gamma(T^{(1,0)}M\otimes T^{(0,1)}M)$ so that their components are  
$g^{i\b j}\p_i\otimes\p_{\b j}, h^{i\b j}\p_i\otimes\p_{\b j}$.  
Then one can define the symmetric bilinear operation \cite{lmz}, \cite{zeit2}:
\begin{eqnarray}
[[,]]:\Gamma(T^{(1,0)}M\otimes T^{(0,1)}M)\otimes \Gamma(T^{(1,0)}M\otimes T^{(0,1)}M)\to \Gamma(T^{(1,0)}M\otimes T^{(0,1)}M)
\end{eqnarray}
written in components as follows: $
[[g,h]]^{k\b l}\equiv (g^{i\b j}\p_i\p_{\b j}h^{k\b l}+h^{i\b j}\p_i\p_{\b j}g^{k\b l}-\p_ig^{k\b j}\p_{\b j}h^{i\b l}-
\p_ih^{k\b j}\p_{\b j}g^{i\b l})$
and looks much less complicated in the jet notation (see section 2). Namely, if $\tilde{\xi},\tilde{\eta}\in J^{\infty}(\mathcal{O}(T^{(1,0)}M)\otimes J^{\infty}(\mathcal{O}(T^{(0,1)}M)$, so that $\xi=\sum_I v^I\otimes\b{v}^I$, $\eta=\sum_J w^J\otimes\b{w}^J$, where $v^I, w^J\in J^{\infty}(\mathcal{O}(T^{(1,0)}M)$, $\b{v}^I, \b{w}^J\in J^{\infty}(\mathcal{O}(T^{(0,1)}M)$, then $
[[\xi,\eta]]=\sum_{I,J}[v^I,w^J]\otimes[\b v^I,\b w^J].$ 

As noted in \cite{lmz},\cite{zeit2}, if bilinear tensor 
$g$ is such that one can associate a  
K\"ahler metrics to it, then the Ricci tensor $R^{i\b j}$ associated with such metric tensor is proportional to 
$[[g,g]]$, more precisely $R^{i\b j}(g)=\frac{1}{2}[[g,g]]^{i\b j}$.

If the complex manifold $M$ has a volume form 
$\Omega$, such that in local coordinates 
$\Omega=e^f dX^1\dots\wedge dX^n\wedge dX^{\bar 1}\wedge\dots dX^{\bar n}$. 
Let us denote the volume form which determines the differential $\mathcal{Q}$ as $\Omega'$, so that $f=-2\Phi_0'$, then $\Phi_0'$ has to be locally a sum of holomorphic and antiholomorphic functions, i.e. it satisfies equation  
$\p_i{\p}_{\b j}\Phi_0'=0$. 

We will refer to the vector field $div_{\Omega}g$ such that 
$(div_{\Omega}g)^{\b j}=\p_ig^{i\b j}+\p_ifg^{i\b j}$, $(div_{\Omega}g)^{i}=\p_{\b j}g^{i\b j}+\p_{\b j}fg^{i\b j}$ as the divergence of bivector field $g$ with respect to the volume form $\Omega$.

Now let the Maurer-Cartan element, closed under $\mathbf{b}^-$, namely the element of $
\Gamma(T^{(1,0)}(M)\otimes T^{(0,1)}(M))\oplus \mathcal{O}(T^{(0,1)}(M)\oplus \mathcal{O}(T^{(1,0)}(M)\oplus
\mathcal{O}_M\oplus \bar{\mathcal{O}}_M$ be defined by its components in the  direct sum, i.e. as $(g, \b v, v, \phi, \b \phi)$.

Then the following Theorem holds, which can be proven by direct calculation.\\

\noindent{\bf Theorem 3.1a.} {\it 
The Maurer-Cartan equation for the differential graded Lie algebra on ${\bf F}^{\cdot}_{sm}\vert_{\mathbf{b^-}=0}$ generated by $\mathcal{Q}$ and $\{\cdot, \cdot\}$ imposes the following system of equations on $g, \phi, \b \phi$ ($\b v, v$ turn out to be auxiliary variables): }

1). {\it Vector field $div_{\Omega}g$, where $\Omega=\Omega'e^{-2\phi+2\b \phi}$ is determined by $f\equiv-2\Phi_0=-2(\Phi_0'+\phi-\b \phi)$ and $\p_i{\p}_{\b j}\Phi_0=0$, is such that its $\Gamma(T^{(1,0)} M)$, $\Gamma(T^{(0,1)} M)$ components are correspondingly holomorphic and antiholomorphic.}

2). {\it Bivector field $g\in \Gamma(T'M\otimes T''M)$ obeys the following equation: 
$
[[g,g]]+\mathcal{L}_{div_{\Omega}(g)}g=0, 
$ 
where $\mathcal{L}_{div_{\Omega}(g)}$ is a Lie derivative with respect to 
the corresponding vector fields.}

3). {\it $div_{\Omega}div_{\Omega}(g)=0$.\\
The infinitesimal symmetries of the Maurer-Cartan equation coincide with the holomorphic coordinate transformations of the volume form and tensor $\{g^{i\b j}\}$. }\\

The constraints 1), 2), 3) coincide with the equations studied in \cite{zeit2},  where it was shown that they are equivalent to Einstein equations, i.e. the following statement is valid.\\ 

\noindent{\bf Theorem 3.1b.} 
{\it If tensor $\{g^{i\b j}\}$ parametrises Hermitian metric, then 
the conditions 1), 2), 3) on $g$ and $\Phi_0$ from Theorem 3.1a are equivalent to Einstein equations 
\begin{eqnarray}\label{components}
&&R^{\mu\nu}={1\over 4} H^{\mu\lambda\rho}H^{\nu}_{\lambda\rho}-2\nabla^{\mu}
\nabla^{\nu}\Phi,\\
&&\nabla_{\mu}H^{\mu\nu\rho}-2(\nabla_{\lambda}\Phi)H^{\lambda\nu\rho}=0,
\nonumber\\
&&4(\nabla_{\mu}\Phi)^2-4\nabla_{\mu}\nabla^{\mu}\Phi+
R+{1\over 12} H_{\mu\nu\rho}H^{\mu\nu\rho}=0,\nonumber
\end{eqnarray}
where $H=dB$ is a 3-form, so that metric $G$, 2-form $B$ and the dilaton field $\Phi\in \mathcal{C}(M)$ are expressed as follows: 
$G_{i\bar{k}}=g_{i\bar{k}}$, $B_{i\bar{k}}=-g_{i\bar{k}}$, $\Phi=\log\sqrt{g}+\Phi_0$,
$G_{ik}=G_{\b i \b k}=G_{ik}=G_{\b i \b k}=0$,
where by $g$ under the square root we denote the determinant of $\{g_{i\b j}\}$. }\\

Physically, the appearance of extra $\log (g)$ part in the dilaton corresponds to the fact that passing from the first order action involves integration over $p_i, p_{\b j}$-variables, which leads to extra contribution to the dilatonic term.\\

\noindent{\bf 3.2. Main conjecture.} Following the ideas of section 3.1, we want to repeat the construction in the case of the complex $(\mathcal{F}^{\cdot}, \mathcal{Q})$. Namely, we consider its jet version $(\mathcal{F}^{\cdot}_{\infty}, \mathcal{Q})$
and its complex conjugate $(\bar{\mathcal{F}}^{\cdot}_{\infty}, \mathcal{Q})$, so that
${\bf F}^{\cdot}_{ \infty}=\mathcal{F}^{\cdot}_{\infty}\hat{\otimes}\bar {\mathcal{F}}^{\cdot}_{\infty}.$ 

It is the jet version of the complex $({\bf F}^{\cdot}, \mathcal{Q})$, such that e.g. the subspace of degree 1 is as follows:  ${\bf F}^{1}\cong\Gamma(E)\oplus\mathcal{C}(M)\oplus\mathcal{C}(M)$. As in section 3.1, the  divergence operator which determines $\mathcal{Q}$-operator, is based on the volume form, given in the local coordinates as $e^{-2\Phi_0'(X)}dX^1\dots dX^n\wedge dX^{\b 1}\dots dX^{\b n}$, so that $\p_i\p_{\b j}\Phi_0'=0$.

We can give ${\bf F}^{\cdot}$ the structure of the homotopy Leibniz bracket by means of the formula  which is the same as in subsection 3.1, however now we have higher homotopies. 
We also note that ${\bf F}^{\cdot}\vert_{\mathbf{b}^-=0}\equiv {\bf F}^{\cdot}_-$ is invariant under $\{\cdot,\cdot\}$. 
Let us formulate the first part of the main conjecture.\\

\noindent{\bf Conjecture 3.1a.} {\it The structure of homotopy Gerstenhaber algebra on 
${\bf F}^{\cdot}$ can be extended to $G_{\infty}$-algebra, so that the subcomplex 
${\bf F}^{\cdot}_-$ is invariant under $L_{\infty}$ operations. }\\

Let us focus on the subcomplex $(\mathbf{F}^{\cdot}_-, \mathcal{Q})$. The space of Maurer-Cartan elements, i.e. the subspace of the elements of degree 2 is:
$
\mathbf{F}_-^2\cong \Gamma(\mathcal{E}\otimes\bar{\mathcal{E}})\oplus \Gamma(E)\oplus \mathcal{C}(M)\oplus\mathcal{C}(M).
$
The elements of this space are defined by means of the components from the direct sum above, i.e. $\Psi=(\mathbb{M}, \eta, \phi,\bar{\phi})$. We will denote the difference $\phi-\bar{\phi}\equiv\Phi_0''$ and $\Phi_0\equiv\Phi_0'+\Phi_0''$. 
Let us formulate the second part of the main conjecture.
\\

\noindent{\bf Conjecture 3.1b.} {\it Let $\Psi=(\mathbb{M}, \eta, \phi,\bar{\phi})$ be the solution of the generalized Maurer-Cartan (GMC) equation for $L_{\infty}$-algebra on ${\bf F}^{\cdot}_-$, so that 
$
\mathbb{M}=\begin{pmatrix} g & \mu \\ 
\bar{\mu} & b \end{pmatrix}.
$
Then the $\eta$-component is auxiliary and is expressed via $\mathbb{M}$ and $\phi, \bar{\phi}$. If $\{g^{i\b j}\}$ is invertible, then $G, B$ obtained from $\mathbb{M}$ via \rf{GB} together with $\Phi=\Phi_0+\sqrt{g}$, where $g$ is the determinant of $\{g_{i\b j}\}$, satisfy the Einstein equations \rf{components}.}\\
 
The space of infinitesimal symmetry generators of GMC equation, i.e. $\mathbf{F}^1$ is given by  $
\mathbf{F}_-^1\cong\Gamma(E)\oplus \mathcal{C}(M)$,
so that any element can be written in components as $\Lambda=(\xi, f)$.

The third part of the conjecture concerns the question how $\xi, f$ are related to 
$\alpha\in \Gamma(E)$ in the transformation formula $
\mathbb{M}\to \mathbb{M}-D\alpha+ \phi_1(\alpha,\mathbb{M})+\phi_2(\alpha, \mathbb{M},\mathbb{M})
$ 
from section 1. 
First, to justify the statement of Conjecture 3.1b, one can show the following via direct computation. \\

\noindent{\bf Proposition 3.2.} {\it 
Let $\Lambda=(\xi, f)\in \mathbf{F}_-^1$ be the generator of the  infinitesimal transformation of the solution of GMC equation. Then after the substitution   
$\xi=\alpha+\frac{1}{2}\mathbb{M}\cdot\alpha$ (where $\mathbb{M}$ is considered as an element of $End(\Gamma(E))$) the transformation of $\mathbb{M}$-component of the solution coincides with \rf{algsym} up to the second order in $\mathbb{M}$.}\\

We note, that the symmetry generated by $f$-part of $\mathbf{F}^1$ element does not affect metric B-field or dilaton. It is easy to check that on  the level of 0th order in $\mathbb{M}$: the symmetry transformation corresponds to the shift of $\phi $ and $\b \phi$  by $f$. One can check, similar to Proposition 3.2, that this symmetry remains redundant for the first and second order. We claim that these statements are exact, namely the following Conjecture is true.\\

\noindent{\bf Conjecture 3.1c.} {\it Let $\Lambda=(\xi,f)\in \mathbb{F}^1_-$, be the generator of the infinitesimal symmetries of  the GMC equation. The corresponding transformation of $\mathbb{M}$-component of the solution of GMC coincide with the diffeomorphism and B-field transformation if  $\xi=\alpha+\frac{1}{2}\mathbb{M}\cdot\alpha$. 
Under conditions of Conjecture 3.1b these transformations  reproduce infinitesimal diffeomorphism transformations and shifts of B-field by exact 2-form, which are the symmetries of equations \rf{components}.} 

\section*{Acknowledgements} I am grateful to A.N. Fedorova for careful reading of the manuscript and to the referee for useful remarks.

\end{document}